\documentclass[12pt]{article}
\usepackage{pic03}
\usepackage{url}
\usepackage{graphicx} 
\usepackage{floatflt}
\usepackage{ifthen,feynmf,amsmath,amssymb} 
\newlength{\wi}   \wi \textwidth
\newlength{\fwi} \fwi 0.95\wi
\newlength{\gwi} \gwi 0.95\wi
\begin{document}

\newcommand{\vektor}[2]
{{#1 \choose #2}}
\newcommand{\SU}{SU}
\newcommand{\SO}{SO}
\newcommand{\U}{U}

\newcommand{\hc}{\mbox{h.c.}}
\newcommand{\Tr}{\mbox{Tr}}

\newcommand{\PC}{Potenzreihenkorrektur}
\newcommand{\PCen}{Potenzreihenkorrekturen}

\newcommand{\Slash}[1]{\mbox{\it #1\hspace{-0.6em}\slash}}

\newlength{\ziffer}
\newcommand{\0}{\settowidth{\ziffer}{0}\hspace*{\ziffer}}

\newcommand{\TeV}{\,\mbox{Te\kern-0.2exV}}
\newcommand{\GeV}{\,\mbox{Ge\kern-0.2exV}}
\newcommand{\mGeV}{\,\mathrm{Ge\kern-0.2exV}}
\newcommand{\MeV}{\,\mbox{Me\kern-0.2exV}}
\newcommand{\keV}{\,\mbox{ke\kern-0.2exV}}
\newcommand{\eV}{\,\mbox{e\kern-0.2exV}}
\newcommand{\km}{\,\mbox{km}}
\newcommand{\m}{\,\mbox{m}}
\newcommand{\cm}{\,\mbox{cm}}
\newcommand{\mm}{\,\mbox{mm}}
\newcommand{\um}{\,\mbox{$\mathrm\mu$m}}
\newcommand{\fm}{\,\mbox{fm}}
\newcommand{\us}{\,\mbox{$\mathrm\mu$s}}
\newcommand{\hz}{\,\mbox{Hz}}
\newcommand{\ipb}{\mbox{pb}^{-1}}
\newcommand{\pb}{\mbox{pb}}

\newcommand{\bea}{\pagebreak[3]\begin{samepage}\begin{eqnarray}}
\newcommand{\eea}{\end{eqnarray}\end{samepage}\pagebreak[3]}
\newcommand{\beq}{\begin{equation}}
\newcommand{\eeq}{\end{equation}}

\newcommand{\Bildzentriert}[3]
{  \begin{figure}[#2]
   \begin{center}
   \input #1
   \end{center}
   \caption{#3}
   \label{#1}
   \end{figure}
}

\newcommand{\Bild}[3]
{  \begin{figure}[#2]
   \input #1
   \caption{#3}
   \label{#1}
   \end{figure}
}

\newcommand{\C}{{C\hspace{-0.65em}I}}
\newcommand{\R}{{I\hspace{-0.35em}R}}
\newcommand{\N}{{I\hspace{-0.35em}N}}

\newcommand{\sm}{Standardmodell}
\newcommand{\ww}{Wechselwirkung}
\newcommand{\ee}{\ensuremath{e^+e^-}}
\newcommand{\qqbar}{\ensuremath{q\bar{q}}}
\newcommand{\qqbarg}{\ensuremath{q\bar{q}g}}
\newcommand{\bbbar}{\ensuremath{b\bar{b}}}
\newcommand{\ppbar}{\ensuremath{p\bar{p}}}
\newcommand{\as}{\ensuremath{\alpha_s}}
\newcommand{\msbar}{\ensuremath{\overline{\rm MS}}}
\newcommand{\Bmax}{B_{\mathrm{max}}}
\newcommand{\Bmin}{B_{\mathrm{min}}}
\newcommand{\Bsum}{B_{\mathrm{sum}}}
\newcommand{\Bdiff}{B_{\mathrm{diff}}}
\newcommand{\Mhigh}{M^2_{\mathrm{h}}/E^2_{\mathrm{vis}}}
\newcommand{\Mlow}{M^2_{\mathrm{l}}/E^2_{\mathrm{vis}}}
\newcommand{\Mhighp}{{M^{2 }_{(p)\mathrm{h}}}/E^2_{\mathrm{vis}}}
\newcommand{\Mlowp}{{M^{2}_{(p)\mathrm{l}}}/E^2_{\mathrm{vis}}}
\newcommand{\MhighE}{{M^{2}_{(E)\mathrm{h}}}/E^2_{\mathrm{vis}}}
\newcommand{\MlowE}{{M^{2}_{(E)\mathrm{l}}}/E^2_{\mathrm{vis}}}
\newcommand{\Mdiff}{M^2_{\mathrm{diff}}/E^2_{\mathrm{vis}}}
\newcommand{\durham}{{Durham}}
\newcommand{\oasspnlla}{\ensuremath{{\cal O}(\alpha_s^2)\oplus\text{NLLA}}}
\newcommand{\ecm}{E_{\mathrm{cm}}}

\newcommand{\asb}{$\alpha_0$}

\newcommand{\eps}{\varepsilon}
\newcommand{\fig}[1]{Fig.~\ref{#1}}
\newcommand{\tab}{Tab.~\ref}
\newcommand{\gl}[1]{Gl.~(\ref{#1})}
\newcommand{\eq}[1]{Eq.~(\ref{#1})}

\newcommand{\oas}{$\cal O$($\alpha_s^2$)}
\newcommand{\oass}{$\cal O$($\alpha_s^3$)}

\newcommand{\sprime}{{\sc Sprime}}
\newcommand{\pythia}{{\sc Pythia}}
\newcommand{\jetset}{{\sc Jetset}}
\newcommand{\ariadne}{{\sc Ariadne}}
\newcommand{\herwig}{{\sc Herwig}}
\newcommand{\excalibur}{{\sc Excalibur}}
\newcommand{\minuit}{{\sc Minuit}}
\newcommand{\zfitter}{{\sc Zfitter}}
\newcommand{\event}{{\sc Event}}

\newcommand{\lep}{{\sc Lep}}
\newcommand{\tristan}{{\sc Tristan}}
\newcommand{\topaz}{{\sc Topaz}}
\newcommand{\lepII}{{\sc Lep~II}}
\newcommand{\delphi}{{\sc Delphi}}
\newcommand{\alephh}{{\sc Aleph}}
\newcommand{\opal}{{\sc Opal}}
\newcommand{\ldrei}{{\sc L3}}
\newcommand{\sld}{{\sc Sld}}
\newcommand{\slc}{{\sc Slc}}
\newcommand{\cleo}{{\sc Cleo}}
\newcommand{\belle}{{\sc Belle}} 
\newcommand{\delana}{{\sc Delana}}
\newcommand{\dstana}{{\sc Dstana}}
\newcommand{\delsim}{{\sc Delsim}}
\newcommand{\Mini}{{\sc Mini}}
\newcommand{\phwmini}{{\sc PHWMini}}
\newcommand{\cargo}{{\sc Cargo}}
\newcommand{\jade}{{\sc Jade}}

\newcommand{\tsppm}{\hspace{\tabcolsep}$\pm$\hspace{\tabcolsep}}

\newenvironment{scaledlist}[0]{
                          \begin{flushleft}
                          \begin{list}{{$\bullet$}}{\setlength{\itemsep}{2ex plus0.2ex}
                          \setlength{\parsep}{0ex plus0.2ex}
                          \setlength{\labelwidth}{2em}}
                         }
                         {
                          \end{list}
                          \end{flushleft}
                         }
\hyphenation{frag-men-ta-tion func-tion}

\title{\bf QCD at ${\mathbf{e^+e^-}}$ Experiments}
\author{Klaus Hamacher  \\
{\em Fachbereich Physik, Bergische Univ., Gau\ss{}-Stra\ss{}e 20,
42097 Wuppertal}}
\maketitle 

%
%
\begin{figure}[h] 
\begin{center}
%
%
\vspace{4.5cm}
%
%
%
\end{center}
\end{figure}

\baselineskip=14.5pt
\begin{abstract} 
A summary of QCD results obtained at \ee\ experiments in
recent years is given. Emphasis is put on basic QCD tests and 
\as\ measurements  with event shapes.
\end{abstract}
\newpage

\baselineskip=17pt

\section{Introduction}
Experimental studies of strong interaction physics at \ee\ machines 
aiming for basic tests of
Quantum Chromodynamics, QCD, enfold 
the determination of the coupling and the quark masses, the parameters 
of the Lagrangian
,
the verification of the basic vertices of QCD, and
tests of the loop corrections interrelated with 
asymptotic freedom and confinement.
Moreover the experimentalists are faced with a wealth of strong interaction
physics phenomena.

$e^+e^-$ annihilation to hadrons provides the simplest strongly interacting
initial state, a quark anti-quark pair.
Actual measurements 
span from the $\Phi$ resonance
at Da$\Phi$ne to $\sqrt{s}\sim 209 \GeV$ at \lepII.
This talk reviews some basic measurements performed by $e^+e^-$ experiments in
recent years.

Results still come up from the measurements at the $Z$
where the cross--section is large and background free precision 
measurements can be performed. At higher energy the measurements are
complicated by frequent initial state photon radiation, leading to
``$Z$ return'' events, and boson ($W^+W^-$ and $ZZ$) pair production.
Still the LEP data in combination with previous low energy results 
allow detailed tests of the energy evolution of hadronic properties.

Off resonance data from the
$B$ factories are important for studies of charm final states as, in
this case, the charmed hadrons are not 
influenced by  $B$ hadron decay products. 
Moreover radiative Onia decays so far provide the only lucid source 
of purely gluonic final states.

\section{Energy Dependence of the Hadronic Multiplicity}
\begin{figure}[tb]
\wi 0.48\textwidth
\fwi 0.99\wi
\gwi 0.7\wi
\begin{minipage}[t]{\wi}
\includegraphics[width=\fwi,height=\gwi]{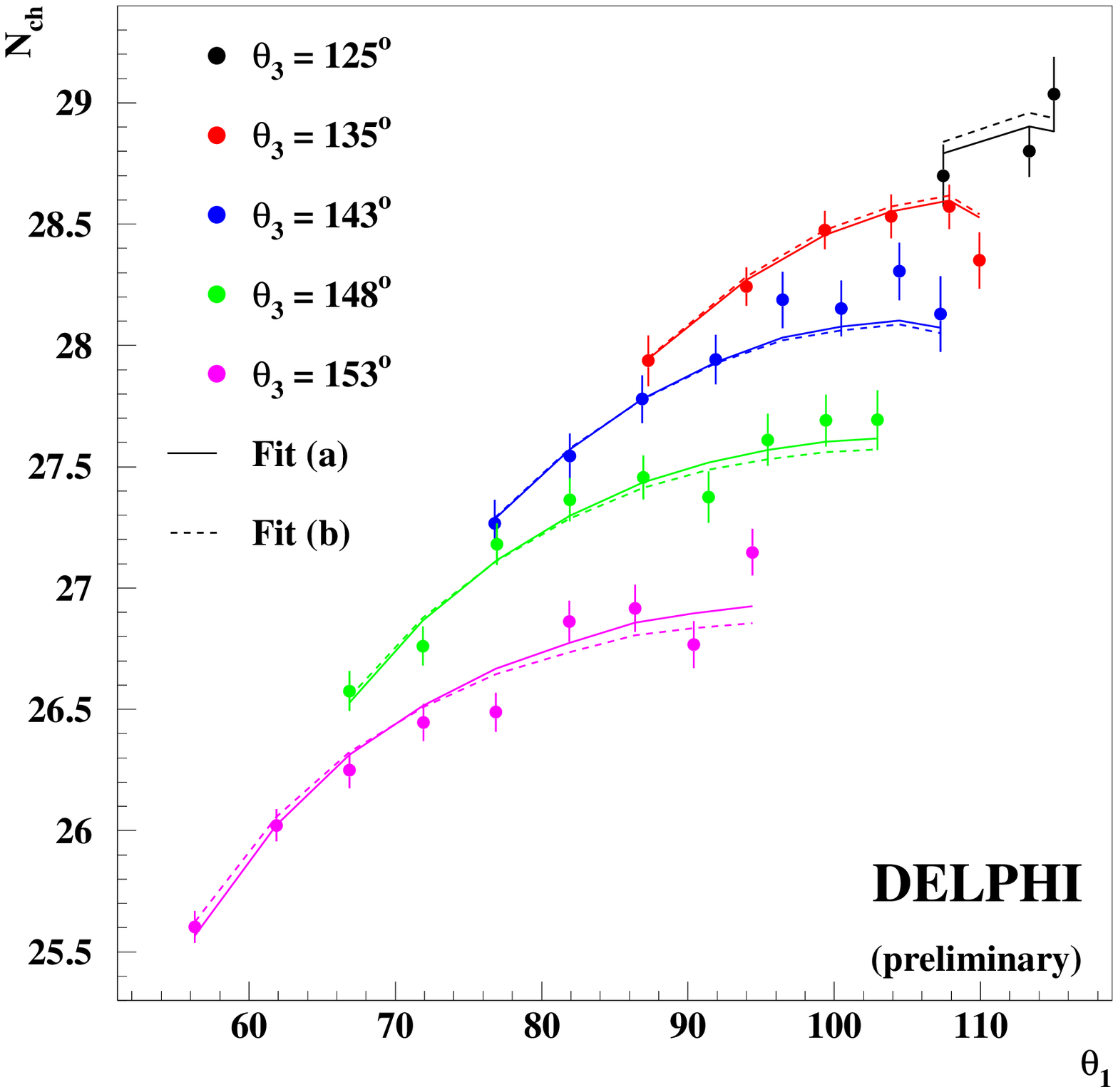}
\vspace*{-.8cm}
\caption{\it Charged hadron multiplicity of three jet events as function of
the inter jet angles.
    \label{f:3jetmul} }
\end{minipage}
\hfill
\begin{minipage}[t]{\wi}
\includegraphics[width=\fwi,height=\gwi]{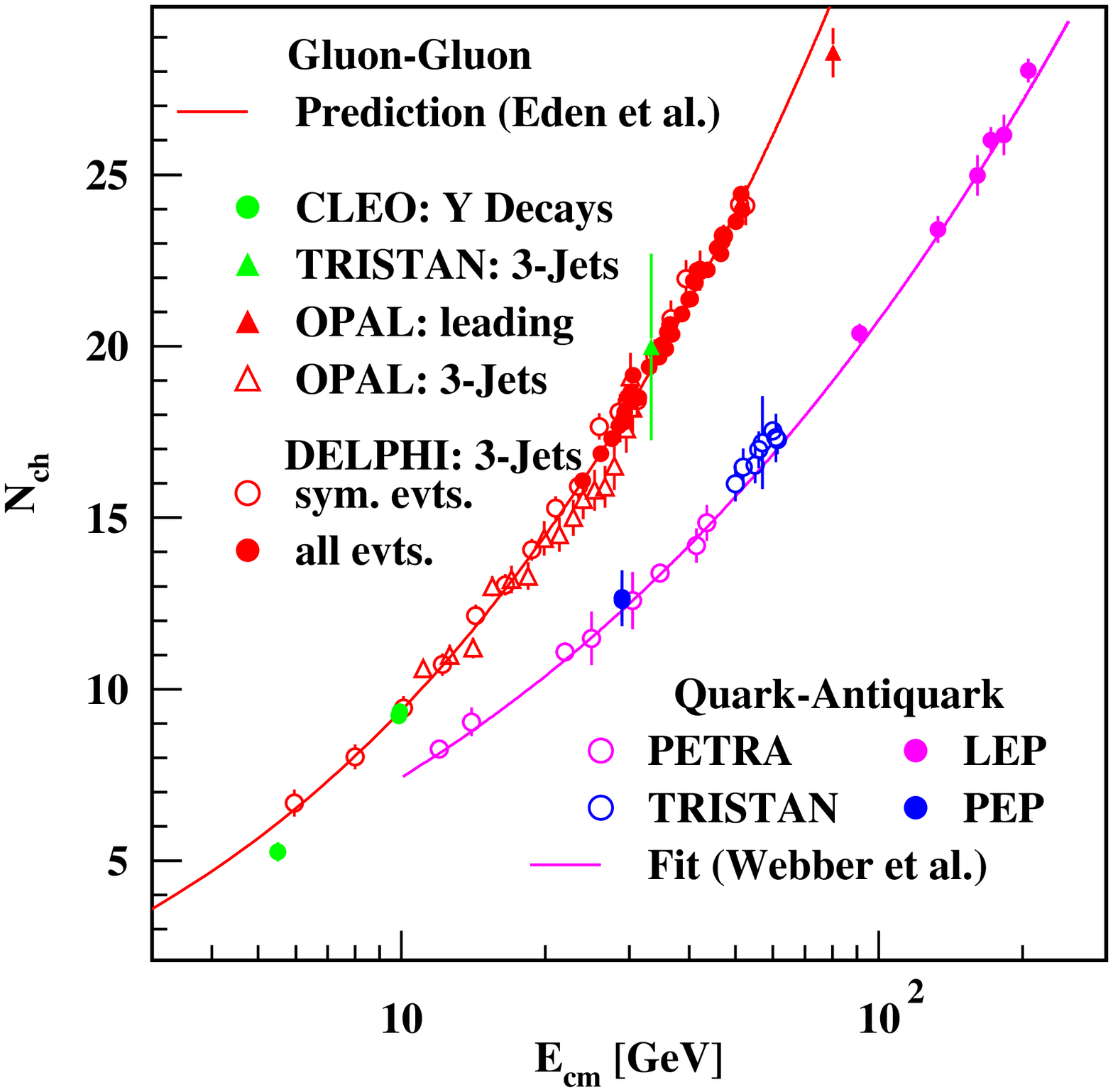}
\vspace*{-.8cm}
\caption{\it Charged multiplicity for \qqbar\ and $gg$ colour singlet states
as function of the CMS energy.
\label{gluemult} }
\end{minipage}
\end{figure}

The dependence of the charged hadron multiplicity on the CMS energy 
up to the highest \lepII\ energies is shown in \fig{gluemult}.
The basic process governing the multiplicity increase is gluon Bremsstrahlung
off the initial \qqbar\ pair ($\propto C_F \cdot \alpha_s$; $C_F=4/3$).
The measurements are well described by parton shower (PS) 
fragmentation models, as well as by  Modified Leading Log Approximation 
(MLLA) and 3NLO calculations \cite{hasko}. 
The calculations intrinsically predict the number
of final state gluons and quarks, the connection to the hadronic multiplicity
being made by Local Parton Hadron Duality (LPHD) assuming
the number of hadrons to be proportional to the number of partons.
The proportionality factor is fitted to data, leaving only the energy increase
predictable.

\begin{figure}[t]
\wi 0.47\textwidth
\fwi 0.99\wi
\gwi 0.6\wi
\begin{minipage}[t]{\wi}
\includegraphics[width=\fwi,height=\gwi]{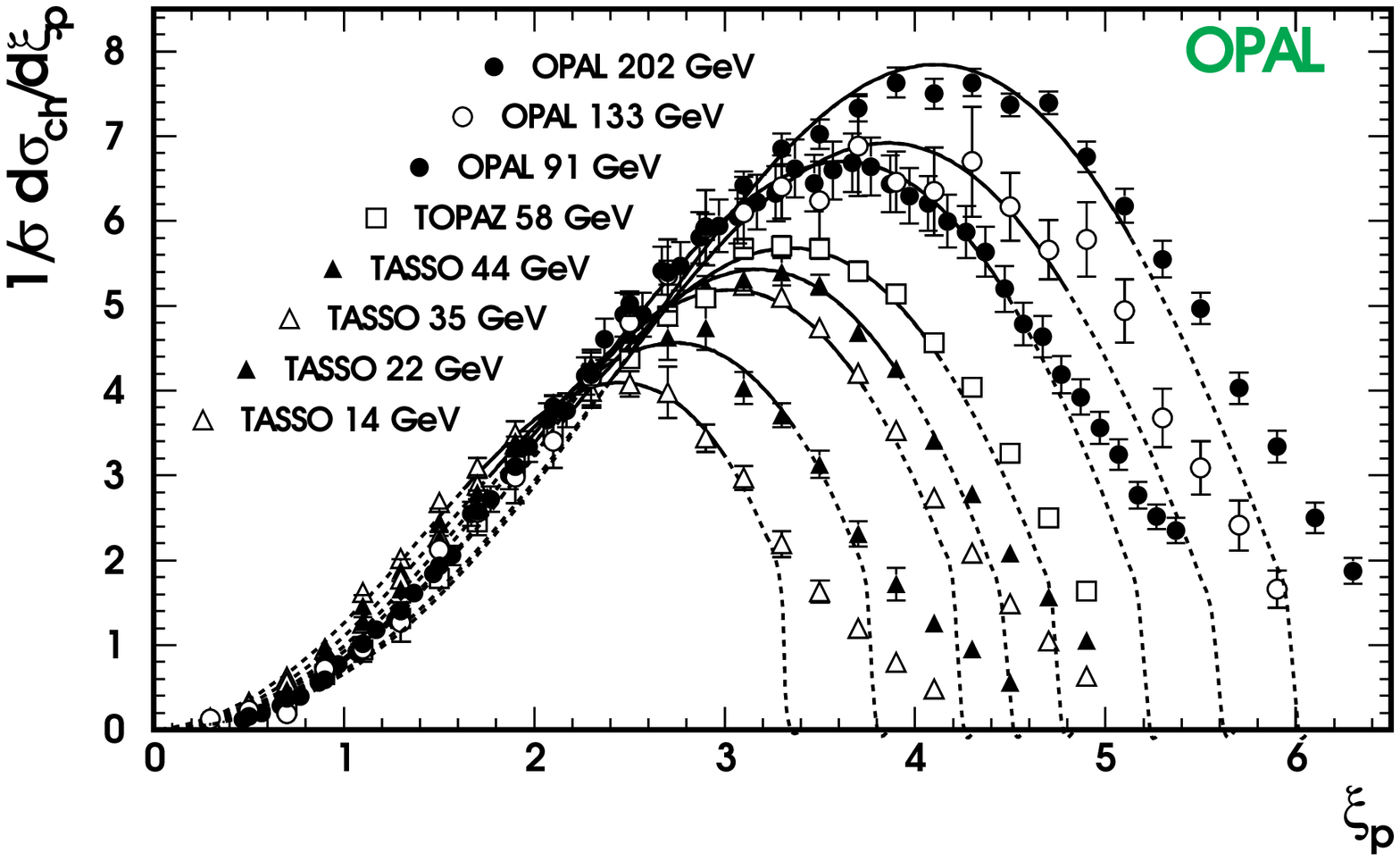}
\vspace*{-.8cm}
\caption{\it
    CMS energy dependence of the spectra of charged hadrons  
    \protect\cite{opalhump}
    as function of 
    $\xi=-\ln x$; $x=2E_h/E_{CM}$.
    \label{hump} }
\end{minipage}
\hfill
\begin{minipage}[t]{\wi}
 \includegraphics[width=\fwi,height=\gwi]{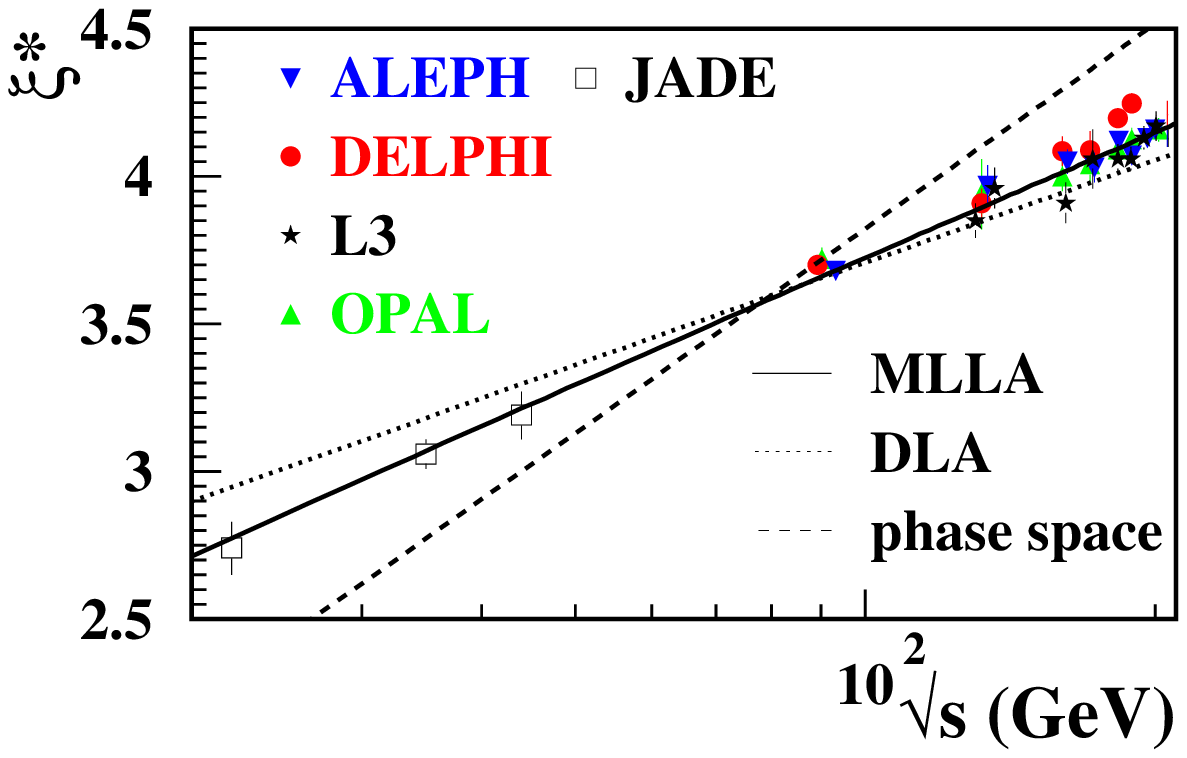}
\vspace*{-.8cm}
\caption{\it
    CMS energy dependence of the maximum of the  $\xi$ distribution, $\xi^*$.
    \label{xistar} }
\end{minipage}
\vspace*{-.4cm}
\end{figure}

An important common aspect of all descriptions is the suppression of soft
gluon emission due to gluon coherence. 
Impressive evidence for this phenomenon comes from the ``hump-backed''
plateau shown in \fig{hump} depicting the suppression of low energy hadrons. 
The data over a wide energy range are compared to the MLLA
``limiting spectrum'' prediction.
The increase of the maximum value, $\xi^*$, of these spectra 
is strongly reduced compared to phase space
expectation (\fig{xistar}).

Independent insight on gluon coherence stems from the multiplicity difference,
$\delta_{bl}$, between light $l=uds$ and $b$ quark events.
In $b$ events the collinear divergence of gluon radiation is regulated by
the $b$ mass. Radiation is suppressed within a ``dead cone''
of opening angle $\Theta\propto m_b/E$, i.e. $\Theta$ decreases towards high
$b$ energies.
The corresponding increase in phase space for radiation 
and gluon coherence lead to the expectation of constant $\delta_{bl}$. 
This expectation has been
recently verified by \lepII\ measurements \cite{deltablexp}. 
By contrast, incoherent radiation would lead to a strong decrease with energy: 
$\delta_{bl}\sim 2\langle N_B\rangle + N_{l\bar{l}}((1-\langle x_B\rangle)^2s)
- N_{q\bar{q}}(s)$, with $\langle x_B\rangle$  the average energy fraction
and $\langle N_B\rangle$ the average decay multiplicity of B hadrons.

Analogously to the quark case the multiplicity in gluon jets  is governed by
gluon Bremsstrahlung off gluons ($\propto C_A\alpha_s$). Due to the higher
colour factor $C_A=3$, a multiplicity increase by the factor $C_A/C_F=2.25$
is expected  at high energy.
Experimental results  obtained mainly in \ee\ 3 jet events 
were found  in the range $1 \dots 1.5$.
The small ratio is now understood considering gluon coherence (leading to
$p_{\perp}$ like scales), finite-energy/non-perturbative (n.p.) 
effects and biases due to jet selection. 

Evidence for n.p. effects comes from the rapidity
spectrum of hadrons in  ``leading'' gluon jets
recoiling with respect to a \bbbar\ system of 
comparably small inter quark angle $\sim
90^{\circ}$ \cite{opalrecoil}.
For small rapidity, i.e. for particles produced first in the hadronisation, 
the production rate is about twice as high as for quark jets of 
corresponding energy. At high rapidity however,  more 
hadrons are produced in quark jets. This is partly due to leading particle 
effects -- quarks are ``valence'' particles of hadrons. Moreover, as
Bremsstrahlung is less effective in quark jets, more energy remains available
and allows high energy hadrons to be formed.
The multiplicity of 3 jet events is predicted \cite{eden} as:
\begin{equation}
N_{\qqbarg}=N_{\qqbar}(s_{\qqbar},y_{cut})+\frac{1}{2}N_{gg}(p_{\perp g}^2)~~~,
\label{3jetmul}
\end{equation}
where $N_{\qqbar}$  is the multiplicity observed in $e^+e^-$ annihilation, 
$N_{gg}$ is the multiplicity for a colour singlet $gg$ pair.
The choice of scales $s_{\qqbar}$ and $p_{\perp g}^2$ is ambiguous. 
It depends on
the part of phase space assigned to the quark and gluon jets in a 3 jet event,
respectively. 
The $y_{cut}$ dependence enters solely due to a bias induced by the 
 jet assignment. 
For a 3 jet event resolved at a selected $y_{cut}^0$  it is prohibited to have 
further gluons emitted at higher $y_{cut}^1 > y_{cut}^0 $ 
as these would lead to $y_{cut}^0=y_{cut}^1$.
Consequently a negative bias on the multiplicity is implied.

The energy slopes of the multiplicity of gluons and quarks are
related by: 
\begin{xalignat}{5}
\left. 
\frac{d{N_{gg}}}{dL'} 
\right
|_{L'=L+c_g-c_q}  
&=  
\frac{C_A}{C_F} \left( 1 - \frac{\alpha_0 c_r}{L} \right )
\frac{d}{dL}N_{q\bar{q}}(L)~~~~~
&
L &= \log{\frac{s}{\Lambda^2}}~~~ & c_i &=\text{const.}
\label{qgdgl}
\end{xalignat}
The solution of this differential equation implies a const.~of integration
suited to absorb n.p. $qg$ differences. 
The complete ansatz describes the observed 
multiplicity for 3 jet events
of general (\fig{f:3jetmul}) and symmetric 
topology \cite{martin}.   
The energy scales are determined from the 3 jet topology and
the three possible assignments of the gluon jet are 
considered by appropriate weighting with the 3 jet
matrix element. \eq{3jetmul} is easily solved for $N_{gg}$. 
The observed increase with energy is 
about twice that of  the quark case (\fig{gluemult}) 
and can be  immediately recognized as due
to the higher colour charge of the gluon.
The quark curve is a MLLA fit to the data, the gluon curve is a 
prediction deduced from the \qqbar\ data using \eq{qgdgl}.
The n.p. offset was fixed by the \cleo\ point at 
$\sqrt{s}\sim 10\GeV$
measured from $\chi_B\rightarrow gg$ decays \cite{cleogg}. 
The prediction agrees with the 3 jet data from 
\tristan\  and 
\lep\ \cite{martin}.
From symmetric 3 jet events \delphi\ measured 
the colour factor ratio:
\begin{equation}
\frac{C_A}{C_F}= 
2.221 \pm 0.032_{\text{stat.}} 
      \pm 0.047_{\text{exp.}} 
      \pm 0.058_{\text{hadr.}}
      \pm 0.075_{\text{theo.}}
\end{equation}

\section{Heavy Quark Fragmentation} 
Despite their complex decay patterns heavy hadrons can be 
reconstructed due to their long lifetime. Since heavy quarks are rarely
produced in fragmentation it is possible to determine even the
fragmentation function of primary $B$ and $D$ hadrons.

\begin{figure}[b]
\wi 0.47\textwidth
\fwi 0.9\wi
\gwi 0.66\wi
\begin{minipage}[t]{\wi}
\includegraphics[width=\fwi,height=\gwi]{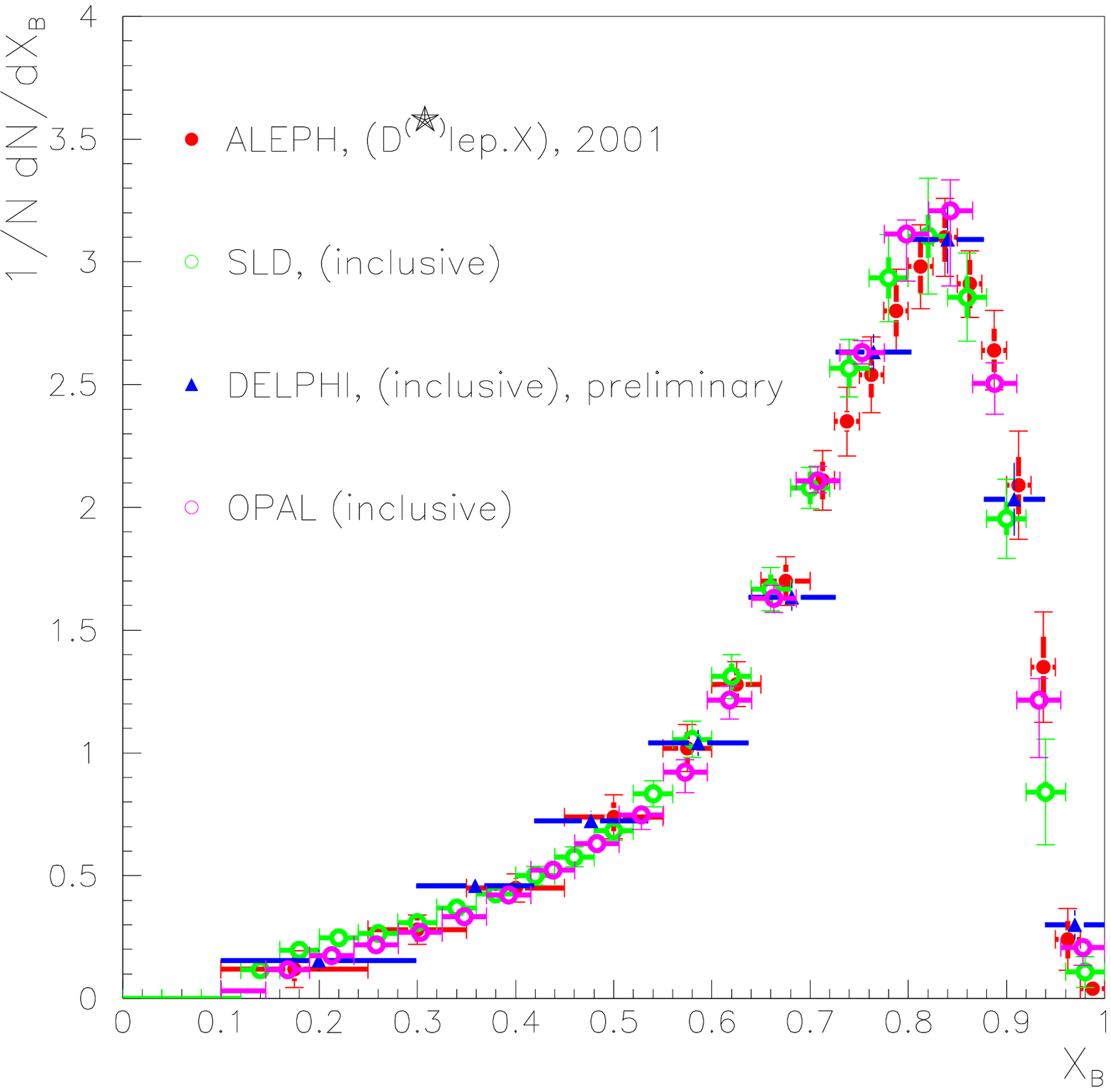}
\vspace*{-.5cm}
\caption{\it B hadron fragmentation function as measured by 
\sld\ and \lep.
\label{fbfragf}}
\end{minipage}
\hfill
\begin{minipage}[t]{\wi}
\includegraphics[width=\fwi,height=\gwi]{xpdstarcleo.eps1}
\vspace*{-.5cm}
\caption{\it Fragmentation function of $D^*$ mesons as measured by \cleo.
    \label{cfragf} }
\end{minipage}
\end{figure}

The expendable micro--vertex detectors of the \lep/\slc\ experiments allow
inclusive $B$ reconstruction with high purity and efficiency
at a ${\cal O}(10\%)$ energy resolution. The new measurements
of the $b$ fragmentation function \cite{bfragf}
(\fig{fbfragf}) agree with each other. Note however, that the data points
are correlated due to the limited energy resolution and the
unfolding applied in the analysis.
The precise average of the scaled energy fraction 
$\langle x_B \rangle =0.715 \pm 0.003$ of weakly decaying $B$'s
is slightly increased compared to previous \ee\ results. This may
be due to the inclusion of $B$ baryons in the sample.
The measurements now allow to distinguish between 
different fragmentation models. 
Consistency is only obtained with the Lund and the similar 
Bowler model while
the data disagree with the widely used Peterson model.
Very precise data on $c$ fragmentation  also became recently
available \cite{cfrag}.
The \belle\ measurement is based on exclusively reconstructed $D^{*+}$
mesons while \cleo\ also uses $D^0$ and $D^+$ decays. 
To avoid contamination from $B$ decays at low $x_D$  off 
resonance data (${\cal L}\sim 3fb^{-1}$) are used. 
For $x_D > 0.5$, where $B$ feed through is kinematically excluded,
also the huge $\Upsilon(4S)$ data sets (${\cal L}\sim 6-25fb^{-1}$) 
are used. 
The momentum resolution is excellent due to the
low CMS energy.
The \belle\ measurement gives
$\langle x_p^{D*}\rangle 
=0.612\pm 0.003_{stat} \pm 0.004_{syst}$ 
and strongly favors the Bowler model.
\cleo\ observes differences in the
spectra of $D^{*+}$, $D^0$ and $D^+$, which are likely due to differing
influence of $D$ resonance decays. Such effects have to be considered
when interpreting the precise results \cite{cfrag}.

The application of the \ee\ heavy quark results
in the description of other processes requires a proper extraction
of the n.p. fragmentation function. The \ee\ analyzers 
now perform this extraction \cite{eliben}.
According to the QCD factorisation
theorem 
the measured function is a convolution of a
pert. piece, describing hard and soft gluon radiation, and the genuine
n.p. fragmentation function.
After Mellin transformation the convolution 
turns into a simple product of transformed contributions. This
equation can be easily solved for the n.p. part and transformed back.
The resulting n.p. $B$ fragmentation function as
evaluated from two \lep\ measurements  agrees well \cite{eliben} .
It is strongly peaked around $x_b \sim 0.9$ 
showing that most of the $b$ quark
energy is conserved in the $B$ hadron.
This function can be directly applied in theoretical predictions provided 
similar assumptions (like renormalisation scheme and factorisation scale) 
are used in the application as in the \ee\ pert. calculation.

The \sld\ experiment  measured the double inclusive
$B$ fragmentation function \cite{slddouble}. 
This function depends on the energies
of and the angle, $\Phi$, between the $B$ particles.
The ratios  $G_{mn}$ of the double to the two corresponding single Mellin 
moments were determined.
These ratios are advantageous for two reasons:
The collinear
singularity is canceled by the finite $b$ mass and 
large log. terms $\propto \ln{E_{cm}/m_b}$ 
cancel to all
orders \cite{brandenburg}. 
The prediction for $G_{mn}(\Phi)$ agrees well with the
measurement\cite{slddouble,brandenburg}. 
As the n.p. fragmentation function enters twice in $G_{mn}$ \sld\
interprets their measurement as a basic test of the QCD factorisation theorem.
The theoretical work \cite{brandenburg} suggests $G_{mn}$ for a precise
extraction of  \as.

\section{Measuring ${\mathbf{m_b(M_Z)}}$ }
Besides the coupling the quark masses are the only free parameters of
the QCD Lagrangian. 
The r\^ole of the masses is similar to that of the coupling.
Besides the pole mass, $M_q$, therefore the renormalised mass, $m_q(Q)$,
is defined, customarily in the \msbar\ scheme. 
Quarks are bound inside hadrons, therefore their masses
can be assessed only via dynamical relations.
At high energy it is possible to determine the mass of the $b$ quark from the
reduction of gluon Bremsstrahlung in $b$ compared to light quark events.
This reduction is 
$\propto m_b^2/p_{\perp g}^2\approx m_b^2/(s\cdot y_{cut})$. The $1/y_{cut}$
factor increases the effect by one order of magnitude compared to the 
directly expected $m_b^2/s$ dependence.
Several coinciding predictions exist (see \cite{listebmasse}).

The $m_b(M_Z)$ measurements performed by most of the \lep/\slc\ experiments 
\cite{listebmasse} proceed as follows:
Jet rates or event shapes are measured for $b$ and light quark
 (or all) events. 
Heavy quark tagging is performed using impact parameter
based techniques.
Detector and hadronisation corrections partly
cancel in the ratio $R_3^{bl}$ of $b$ and light distributions leading
to controllable uncertainties 
despite the weak effect 
$\sim 3\%$ at $\sqrt{s}=M_Z$.

\wi 0.4\textwidth
\fwi 0.99\wi
\gwi 01.1\wi
\begin{floatingfigure}[bh]{\wi}
\flushleft
\begin{minipage}[t]{\wi}
\vspace*{-.2cm}
\includegraphics[width=\fwi,height=\gwi]{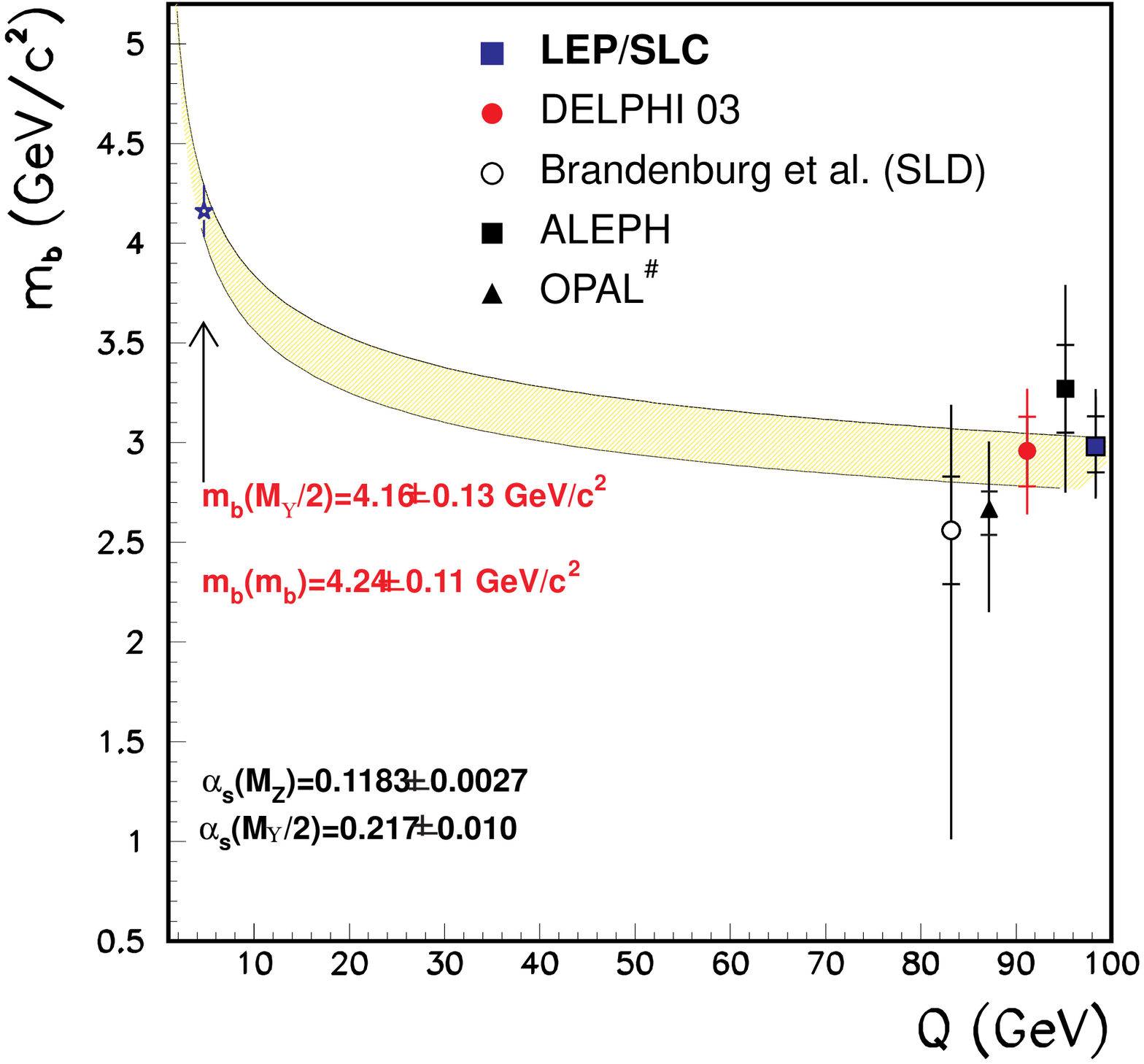}
\caption{\it E dependence of $m_b^{\overline{MS}}$
     }\label{mbrun}
\end{minipage}
\end{floatingfigure}
\noindent
Improved understanding of the hadronisation correction has been gained lately
\cite{mariajose}.
Apparent differences between models 
were traced to an imperfect
prescription of the $x_B$ distribution. The new \delphi\
analysis avoids badly described regions 
henceforth leaving the
associated uncertainty unimportant.
A toy simulation connecting the different $b$ quark masses and effects in
\pythia\ with each other showes for the first time the expected stability 
of the hadronisation correction against parameter changes.
However, the uncertainty of the $b$ mass in the model, which could be 
identified with the pole mass, now appeared as important additional uncertainty.

The expected $\propto y_{cut}^{-1}$ dependence  of $R_3^{bl}$ 
is seen \cite{listebmasse} and
the data are consistent with the NLO expectation for $m_b(M_Z)\sim 3\GeV$.
This small value already indicates the expected ``running'' of the
mass seen in \fig{mbrun}. Here the average $m_b(m_b)$
determined from bound states
is compared to the $m_b(M_Z)$ 
results and their average:
\begin{xalignat}{1}
m_b(M_Z)^{ADO/SLC}&=2.95 \pm 0.15_{stat}\pm 0.24_{hadc/theo}~~~.
\end{xalignat}
For the averaging, theoretical and hadronisation
uncertainties were taken fully correlated. The stat.~error of \opal\
determined from a correlated average has been scaled according to 
PDG rules with $\sqrt{\chi^2/N_{df}}$ of this average.
Improved measurements of $m_b(m_b)$ from moments of the lepton spectra of 
$B$ decays or using lattice calculations  are expected \cite{rainerZ}. 
The fall off of the mass with energy is given by the Renormalisation Group Equation:
\begin{xalignat}{2} 
\frac{\partial \ln m(Q^2)}{\partial \ln Q^2}&=-\gamma_m(\alpha_s)
& \gamma_m ~ \sim ~
\setlength{\unitlength}{1.mm}
\begin{fmffile}{mf_mbloop} 
  \parbox[c]{30mm}  
   {
   \begin{fmfgraph}(28,10) \fmfleft{i} \fmfright{o1}
   \fmfpen{thin}
   \fmf{plain}{i,v1}
   \fmf{plain}{v1,v2}
   \fmf{plain}{v2,o1}
   \fmf{gluon,left=1.,tension=0}{v1,v2}
\end{fmfgraph}}
\end{fmffile}~~~,
\end{xalignat}
where $\gamma_m$ is the mass anomalous dimension, sensitive to loop corrections
to the quark propagator. The succesful description of the $b$ quark mass
running implies a basic test of the quantum loop structure of QCD.

\section{Four Jet Events and Colour Factors}
In 4 jet events all basic QCD vertices
except the suppressed quartic gluon vertex occur as building blocks already at
tree level. 
At tree level the 4 jet cross--section can be decomposed into terms 
proportional to the colour factor products 
$C_F^2$, $C_F C_A$ and $C_F n_f T_F$.
This possibility is due to the general helicity structure of the
process and allows measurements of 
the colour factor ratios $C_A/C_F$ and $n_f T_F/C_F$
from the 4 jet angular distributions. This is equivalent to a
determination of the gauge group underlying the strong interaction.
Recently NLO predictions for the cross--section became available
(for references see \cite{opalaleph4jet})
improving the data description. At the cost of a leading order determination 
it is thus possible to determine directly the colour factors $C_A$ and $C_F$,
as well as \as\
in combination with measurements of the 3 jet rate \cite{opalaleph4jet}.
The current averages of the colour factors are:
\begin{xalignat}{3}
C_F &= 1.34 \pm 0.23 & C_F &= 2.97 \pm 0.5
\end{xalignat}
The results for the colour factor ratios are included in \fig{wim}.
A drawback of all measurements is the residual dependence on a necessary
hadronisation correction. Here QCD is presumed uncared if MC
models or QCD power corrections are used.

\section{${\mathbf{\alpha_s}}$ from Event Shapes}
The measurement of \as\ with event shape observables probes the amount of
gluon radiation in the hadronic final state $\propto C_F\cdot\alpha_s$. 
It is insensitive to
the underlying electroweak physics. 
Event shape observables are required to be insensitive to
soft and collinear gluon emission as well as to hadronisation effects.
A typical observable is ${t=1-Thrust}$. 
In the following an arbitrary observable is denoted by $y$, where low values 
indicate small and large values ($y \sim 0.3$) imply hard gluon radiation.

The \as\ measurement proceed as follows: The experimental distributions
are determined and corrected for cuts, limited detector acceptance and
resolution. At high energy $Z$ return events and two boson production need
special regard.

Two different types of predictions are available based on NLO
\as\ matrix element (ME) and Next to LLA (NLLA) calculations.
The former applies to hard radiation, the latter to the two jet
regime (small $y$) where multiple gluon emission needs to be resummed.
Both calculations can be matched in order to obtain a prediction valid for
a wide region of phase space.

When comparing theory to data it is important to correct the influence of the 
n.p. hadronisation. The classical correction is given by
the ratio of the $y$ distribution on parton (that is just before hadrons are
formed) to hadron level in a Monte Carlo  model. 
Alternatively so called Power corrections (PC) are used.

\subsection{{ \as} from NLO Matrix Elements}
\begin{figure}[hb]
\wi 0.47\textwidth
\fwi 0.99\wi
\gwi 0.9\wi
\begin{minipage}[tbh]{\wi}
\includegraphics[width=\gwi,height=\gwi]{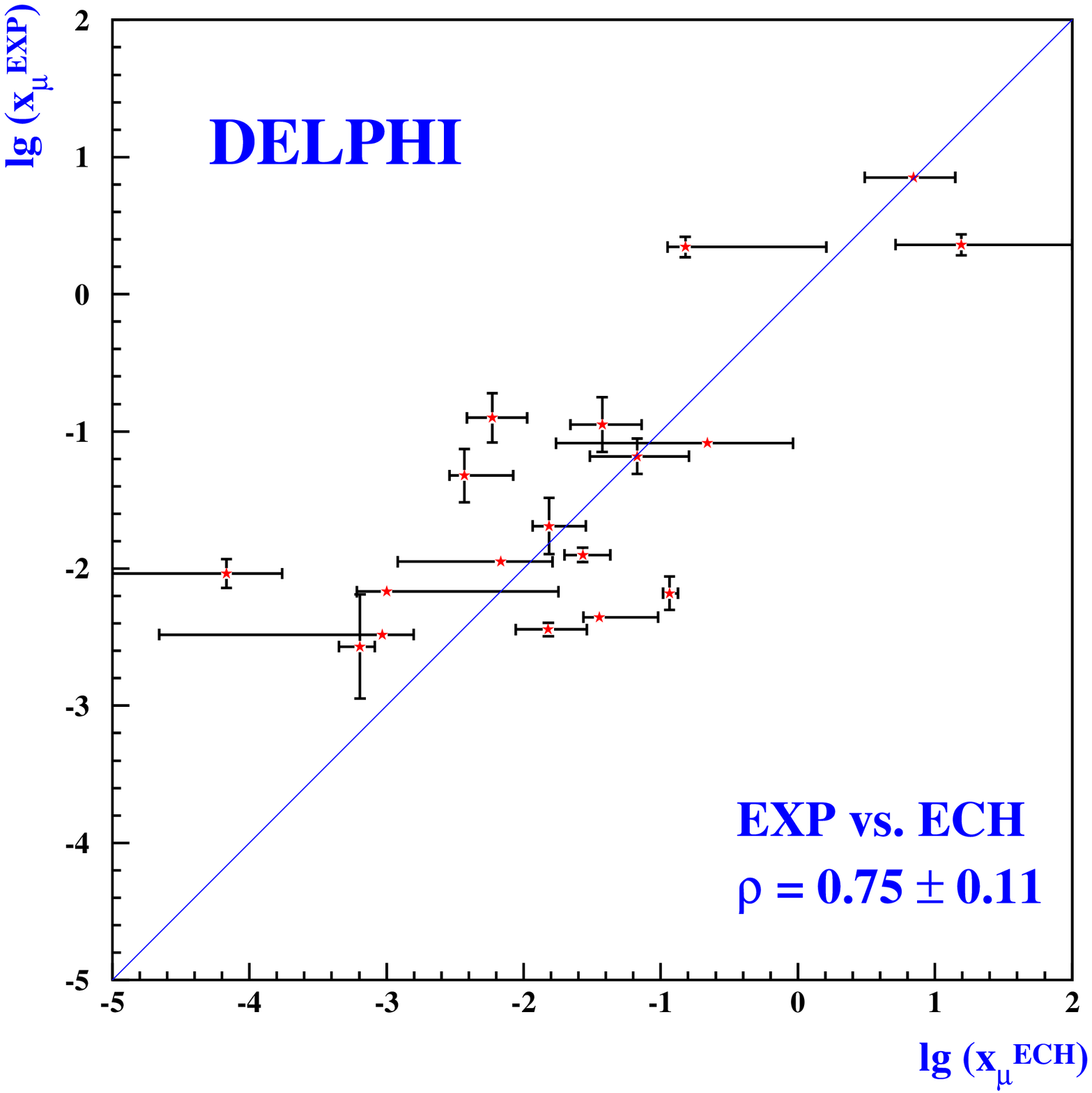}
\vspace*{-.4cm}
\caption{\it Correlation between the experimentally optimised scale 
and the scale corresponding to the ECH scheme.
    \label{correlplot} }
\end{minipage}
\hfill
\begin{minipage}[tbh]{\wi}
\vspace*{-.4cm}
\includegraphics[width=\fwi,height=\gwi]{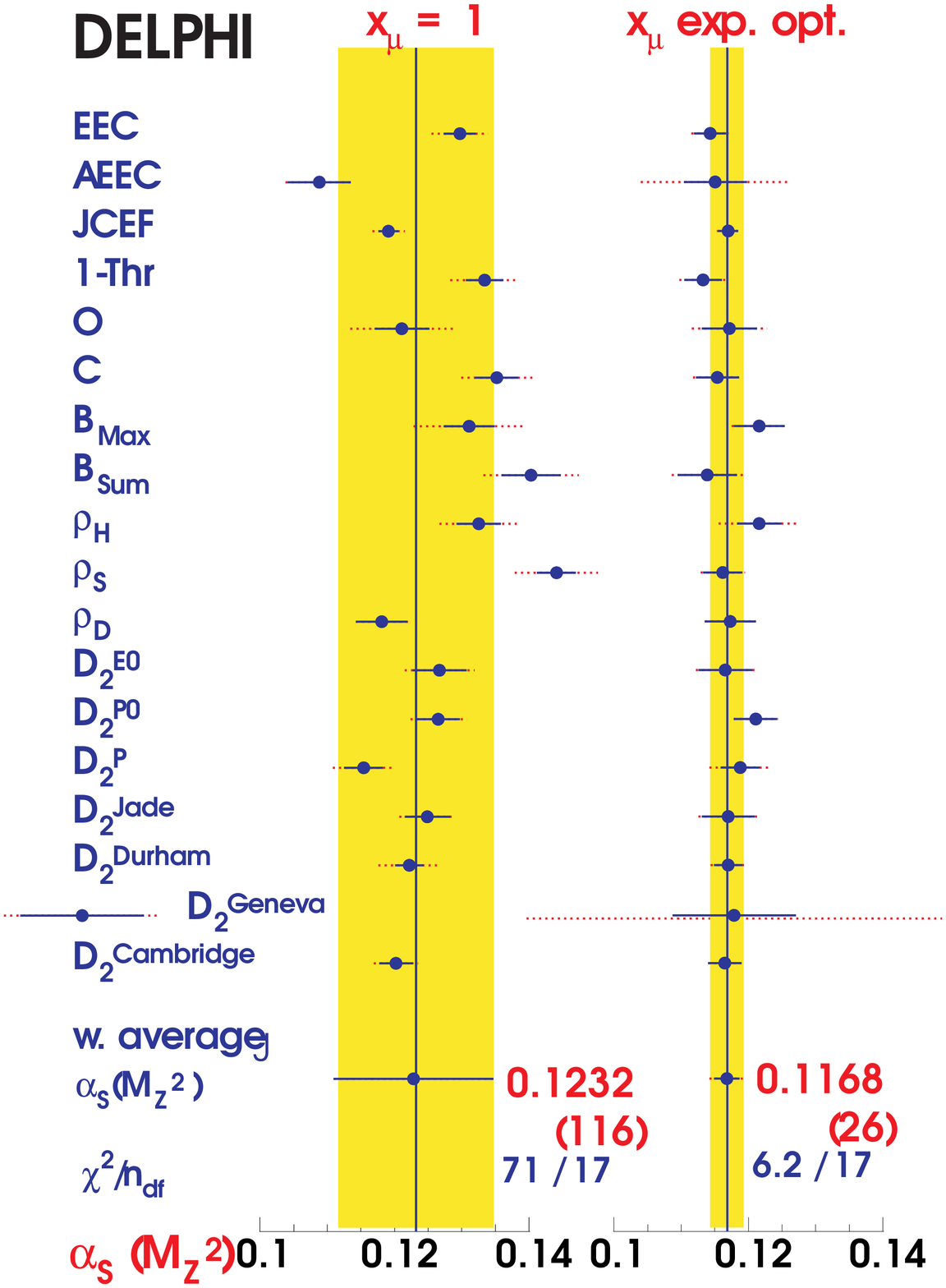}
\vspace*{-.8cm}
\caption{\it $\alpha_s(M_Z)$ as determined from 18 observables from \delphi\
data using ${\cal O}(\as^2)$ theory.
    \label{asdelphi} }
\end{minipage}
\end{figure}
\noindent
The NLO prediction for the distribution of a 3 jet like observable is:
\begin{xalignat}{3}
D(y) &= \frac{1}{\sigma}\frac{d\sigma}{dy}= 
A(y) \bar{\alpha}_s(\mu) + 
[A(y) \beta_0 \ln x_{\mu} + B(y)]\bar{\alpha}^2_s(\mu)  &
\bar{\alpha}&=\frac{\as}{\pi}
\label{secorder}
\end{xalignat}
where $A(y)$ and $B(y)$ are pert.~calculated coefficients. 
For small $y$ this prediction receives important higher order corrections
$\propto \as\ln^2 y$.
Uncalculated higher oder terms lead to the dependence on the unknown 
(observable dependent) renormalisation scale $\mu$ ($x_{\mu}=\mu^2/s$).
The $x_{\mu}$ term changes the slope of $D(y)$, typically falling off
exponentially at not too small $y$.
Satisfactory agreement ($\chi^2/N_{df}\sim 1$) of the prediction is
only obtained with optimised $x_{\mu}$.
Taking $x_{\mu}=1$, that is the pure \msbar\ prediction, will in general 
lead to a dependence of \as\ on the chosen fit range.
This fact was long known for small $y$  but has also been
verified at high $y$ with the precise $Z$ data \cite{siggi}.
In fact this behavior must persist at high $y$ as a consequence of the
normalisation of $D(y)$ to the total cross--section (see \eq{secorder}).

The optimised scales for observables with large positive
(negative) second order prediction tend to be small (high).
Currently it is little understood why the experimentally optimised scales
correlate with the theoretically motivated FAC/PMS/ECH scale
choices (\fig{correlplot} \cite{siggi}).
The \as\ results obtained using optimised scales only show a scatter of about
$2 \%$.
For the $x_{\mu}=1$ results, however, due to the aforementioned discrepancies
the scatter is bigger $\sim 10\%$ for the same $y$ intervals in the \as\ fits.
In this case an additional theory uncertainty is required in order to assure 
consistency of the \as\ results obtained from different observables.

\subsection{{\as} from Matched \oasspnlla\ Predictions}
The NLLA prediction 
has the general form:
\begin{xalignat}{3}
R(y) = \int_0^y D(y)dy &= \exp \left \{ Lg_1(\as L) + g_2(\as L) \right \}
& L&=-\ln y~~~.
\label{nlla}
\end{xalignat}
The function $R(y)$, however,  does not obey the phase space boundary conditions
$R(y_{max})=1$, $\partial_y R(y_{max})=0$. This imperfection is corrected by the
transformation:
\begin{equation}
L \longrightarrow L^{\prime} = \frac{1}{p} \ln 
\left [ \frac{1}{(x_L y)^p} - \frac{1}{(x_L y_{max})^p} + 1 \right ] ~~~,
\end{equation}
where $x_L$ and $p$ are arbitrary parameters \cite{lepwgtheo}. 
This transformation, though
primarily modifying the prediction at high $y$ also implies a \% level shift 
in the \as\ fit range.

In order to obtain a prediction valid over a large range of phase space the
NLLA and NLO ME predictions for $x_{\mu}=1$ are merged. 
Duplicated terms are subtracted
by expanding the NLLA prediction in a  power series in \as. This subtraction
is ambiguous as it can be performed for $R$ or
 $\ln{R}$ (see \eq{nlla}), respectively. 
\begin{figure}[htb]
\wi 0.47\textwidth
\fwi 0.99\wi
\gwi 0.9\fwi
\begin{minipage}[t]{\wi}
\includegraphics[width=\fwi,height=\gwi]{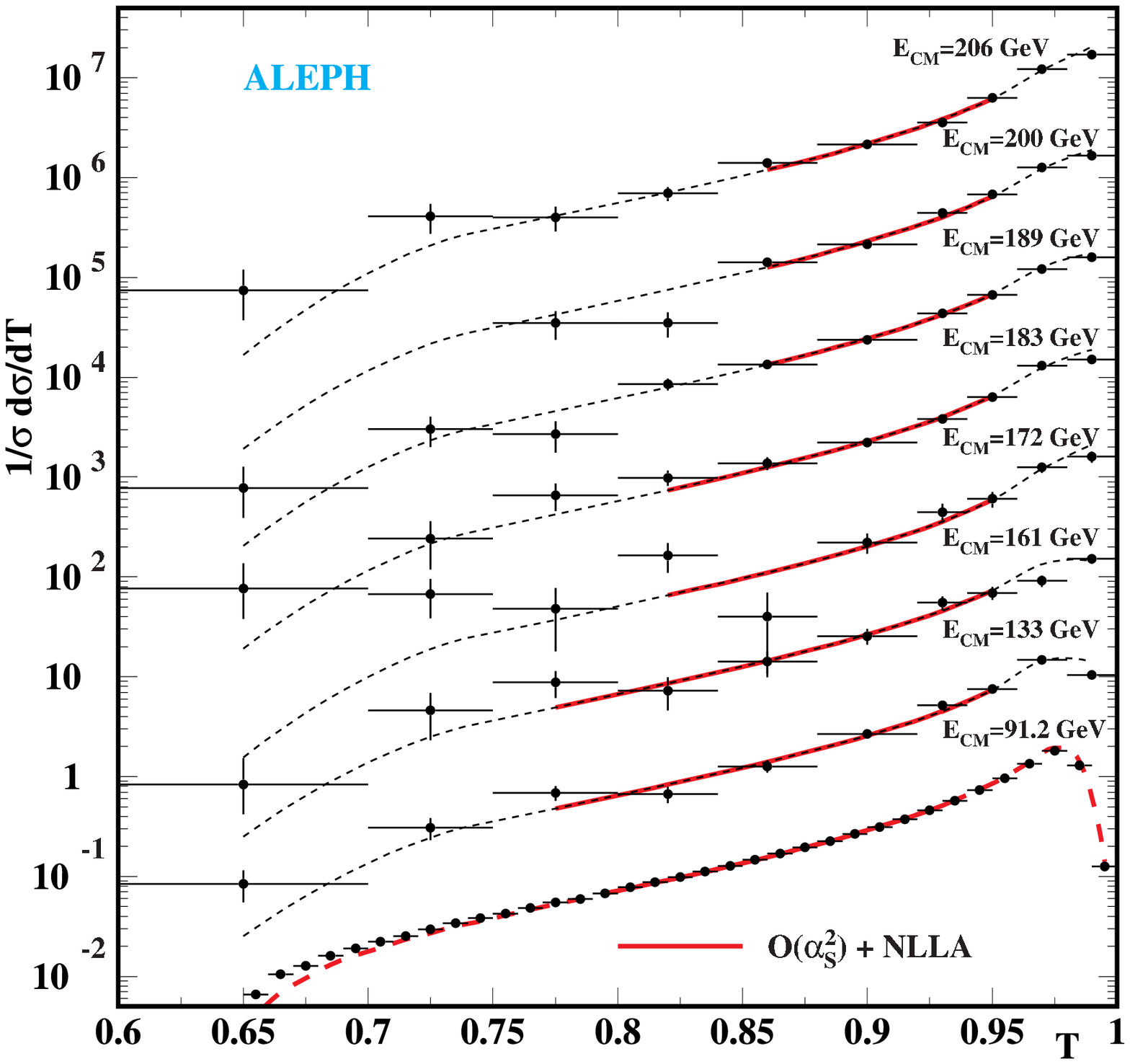}
\vspace*{-.8cm}
\caption{\it Energy dependence of the $Thrust$ distribution compared to
the matched \oasspnlla\ prediction.
    \label{Ethrust} }
\end{minipage} 
\hfill
\begin{minipage}[t]{\wi}
\includegraphics[width=\fwi,height=\gwi]{asrunninglep.eps1}
\vspace*{-.8cm}
\caption{\it Energy dependence of \as\ as determined by the \lep\ QCD W.G.
For data references see \cite{lepwgcomb}.
    \label{aslep} }
\end{minipage}
\end{figure} 

In \fig{Ethrust} a comparison of the $T$ distributions measured between the
$Z$ and 206\GeV\ and the matched 
prediction is shown.
The data are well described over the full range of energy and $T$. 
Only for the precise 
$Z$ data a slight difference in the slope is indicated. This difference is 
weaker for most other observables.

The \lep\ QCD Working Group attempts to combine the \as\ measurements from the
 high energy ( low statistics ) \lepII\ measurements. 
The \oasspnlla\ prediction is use for the observables 
$1-T$, $M_h$, $C$, $B_t$, $B_w$ and $y_3$. The
theory implementations and hadronisation corrections have been compared, the
treatment of correlations and the theory uncertainty was discussed. 
It was seen that often badly described observables
dominate in the combination and a downward bias in \as\ was observed. 
This bias is due to the
proportionality of the \as\ theory uncertainty to $\as^3$. 
To avoid these faults a
novel definition of the theory uncertainty, the uncertainty band method, was
introduced: the matched reference prediction with a 
fixed initial $\as(M_Z)$ was varied by changing the renormalisation scale
($1/2 < x_{\mu} < 2$), not applying or  applying the NLLA
phase space condition with $2/3 < x_{L} < 3/2$, $p=1;2$ and
changing from $\ln{R}$ to $R$ matching. 
The envelope of the variations defines the
uncertainty band  which in turn determines the 
uncertainty of the 
reference \as. Using this fully correlated uncertainty as well
as the experimental and hadronisation uncertainties the \lep\ average \as\ 
is then calculated. 
The procedure is iterated until the result is stable.
The results for $\as(M_Z)$ are \cite{lepwgcomb}: 
\begin{xalignat}{1}
\text{LEP I~:~}\qquad & 
0.1200 \pm 0.0002_{stat} \pm 0.0008_{exp} \pm 0.0010_{had} \pm 0.0048_{th}\\
\text{LEP II:~}\qquad & 
0.1201 \pm 0.0005_{stat} \pm 0.0010_{exp} \pm 0.0007_{had} \pm 0.0045_{th}
\end{xalignat}
The observed energy dependence of \as\ shown in \fig{aslep} agrees well with
the expectation. Despite the strongly different statistics the overall 
uncertainties of the \lep\ I and II results are
comparable.
This is due to the dominance of the theory uncertainty and a decrease 
in the hadronisation ($\propto 1/E$)
and theory uncertainties ($\propto\as^3(E)$) with energy.
The total uncertainty of the order 4\% is confirmed by the scatter of the 
\as\ results from the six individual observables studied.

\subsection{Power Corrections (PC)}
Assuming for the hadronisation process a longitudinal,
``tube--like'' phase space with fixed average transverse
momentum, where only the rapidity range expands with increasing energy, $E$, 
it is easily shown that hadronisation corrections vanish 
$\propto 1/E$. 
Also the replacement of the unknown analytical behavior of the strong
coupling below an IR matching scale $\mu_I\sim 1-2 \GeV$ by an 
effective average value
$\alpha_0(\mu_I)$ leads to a inverse power law behavior \cite{dw}:
\begin{equation}
{\cal P}  =    \frac{4C_F}{\pi^2}{\cal M}
\frac{\mu_I}{\ecm} 
  \left[{{\alpha}_0(\mu_I)} - \alpha_s(\mu)
        - \left(b_0 \cdot \log{\frac{\mu^2}{\mu_I^2}} + 
\frac{K}{2\pi} + 2b_0 \right) \alpha_s^2(\mu) 
\right]\label{DWpower}
\end{equation}
For most observables the value $y$ of the pert.~calculation is just shifted 
by $+c_y{\cal P}$. Correspondingly 
its mean value $\langle y \rangle$ gains an additive  correction.
The  constants $c_y$ have been evaluated for NLLA resummable 
observables. 
The overall size of the 
PC, i.e. $\alpha_0$, is determined from a fit to data. 
Within the uncertainty of the calculation ($\sim 20\%$) universal $\alpha_0$
(and \as) values are expected from the different observables.

The model scetched above is able to succesfully describe the data on
distributions and means over a wide energy range 
\cite{ralle,powerrescollection}.
However, the obtained  values of $\alpha_0\sim 0.5$
are only consistent within the
quoted uncertainty of the model. Specifically 
$\alpha_0$ is smaller for mean values than for distributions. 
Due to the obvious correlation of $\alpha_0$ and \as\ (see \eq{DWpower})
this leads to a corresponding difference in $\alpha_s(M_Z)$. The \jade\
experiment for instance reports \cite{powerrescollection}:
\begin{xalignat}{3}
&\text{means:}\quad &  
\alpha_s(M_Z) & =0.1187  \pm0.0014_\mathrm{fit} 
\pm0.0001_\mathrm{sys}{^{+0.0028} _{-0.0015}}_\mathrm{th}\\
&\text{distributions:}\quad &  
\alpha_s(M_Z) & =0.1126  \pm0.0005_\mathrm{fit} 
\pm0.0037_\mathrm{sys}{^{+0.084} _{-0.060}}_\mathrm{th}
\end{xalignat}
This discrepancy persists even if hadron mass induced PC's are 
included\cite{ralle}.

\subsection{Overview of \as\ Results from LEP}
\begin{figure}[t]
\wi 0.47\textwidth
\fwi 0.99\wi
\gwi 0.9\wi
\begin{minipage}[t]{\wi}
\includegraphics[width=\fwi,height=\gwi]{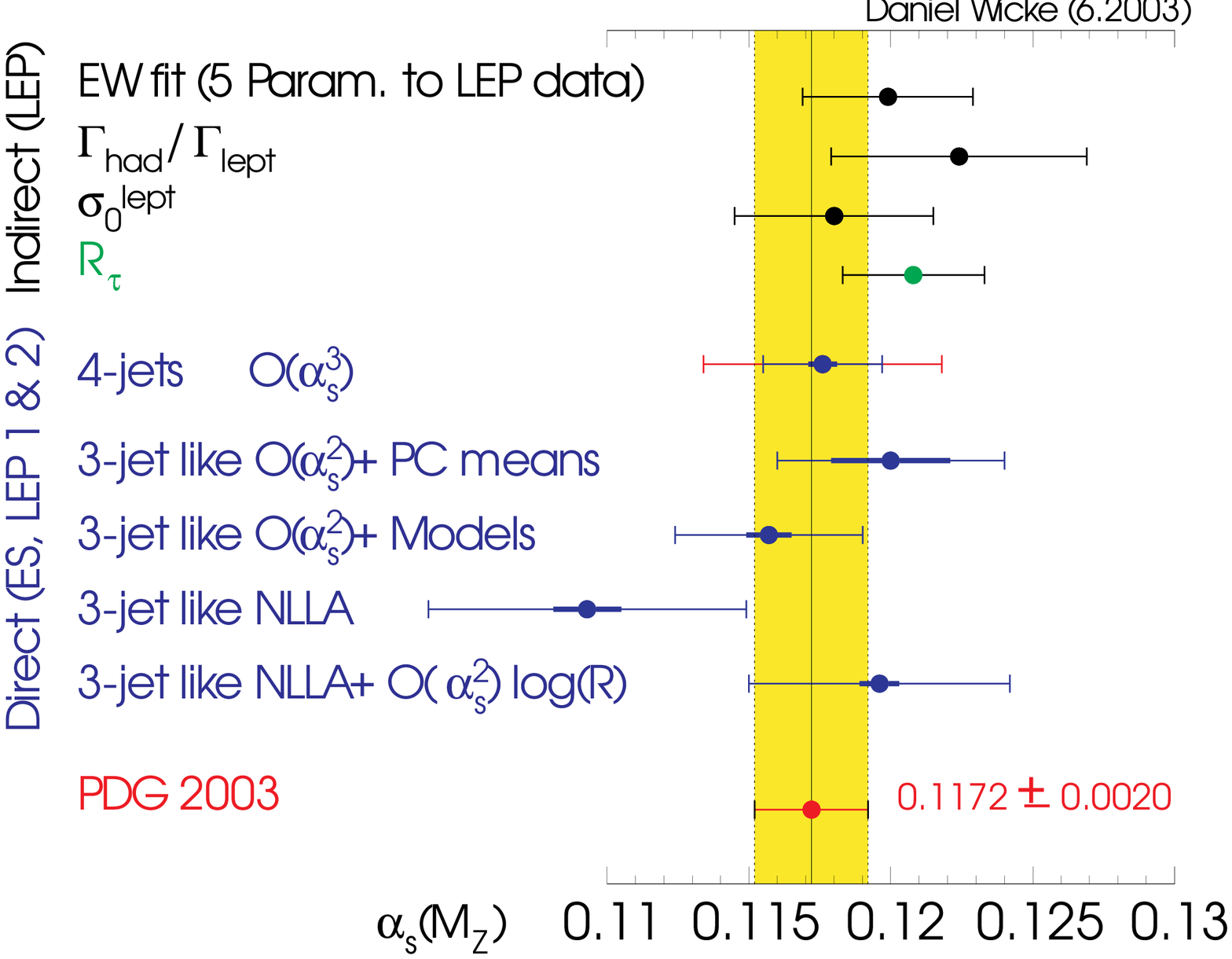}
\vspace*{-.8cm}
\caption{\it Comparison of some important \as\ results from \lep.
    \label{alphascomp} }
\end{minipage}
\hfill
\begin{minipage}[t]{\wi}
\includegraphics[width=\fwi,height=\gwi]{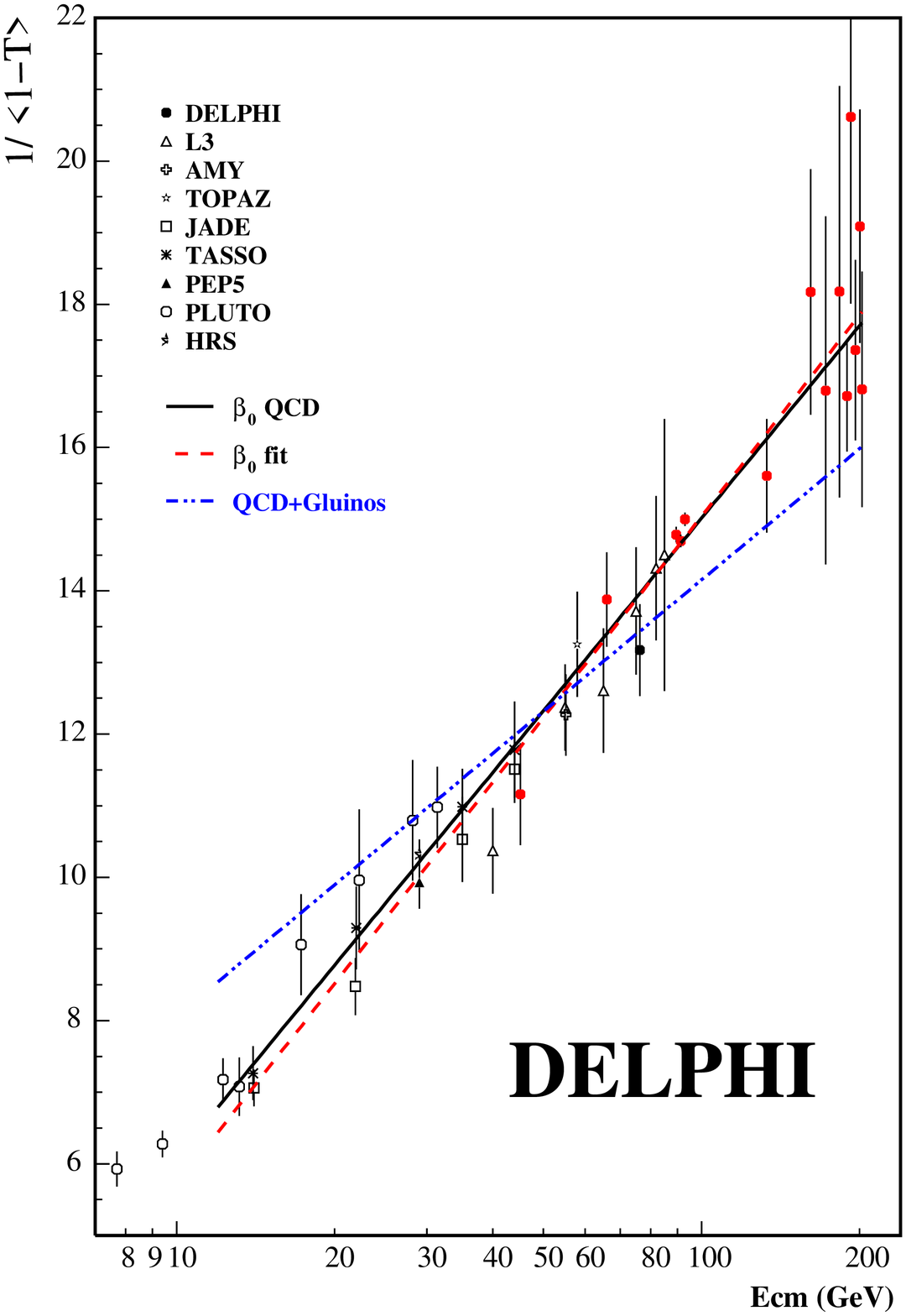}
\vspace*{-.8cm}
\caption{\it Energy evolution of $\langle 1-T \rangle$. 
The slope determines the $\beta$ function.}
    \label{betaplot}
\end{minipage}
\vspace*{-.3cm}
\end{figure}

\fig{alphascomp} compares some important \as\ results obtained from \lep\ data.
All results are compatible with the current PDG average.
The indirect measurements from the $Z$ shape and $R_{\tau}$ are slightly 
above the average. The scatter of these partly highly correlated results 
obtained using NNLO theory confirms their error estimates.

The classical ( 3 \& 4 jet ) NLO direct measurements (with $x_{\mu}^{opt}$
and Monte Carlo  hadronisation) agree well with the PDG average.
The highly correlated result using matched theory
is slightly higher also in comparison to the pure NLLA result.
A shift towards higher \as\ can be understood if the 
${\cal O}(\as^2)|_{x_{\mu}=1}$
part of the matched prediction underestimates the slope of the
data distribution. 
The results for means using PC's are slightly higher than the PDG average.
Overall the scatter of the direct results confirms an error
of $3-4\%$ of \as\ typically assigned to the individual measurements.
Improved predictions (like complete ${\cal O}(\as^3)$ ME's) are 
mandarory for any significant reduction of the uncertainty.

\subsection{The Running of {\as}}
The striking strong interaction physics phenomena of asymptotic freedom and
confinement are intimately coupled to the energy evolution, the 
``running'' of \as. In QCD this is controlled by the Renormalisation Group
Equation (RGE):
\begin{xalignat}{3}
\frac{\partial \alpha_s}{\partial \ln Q^2} &= 
-\alpha_s^2\cdot \beta(\alpha_s)\quad
&
\frac{\partial \alpha_s^{-1}}{\partial \ln Q^2} &=\beta(\alpha_s)=\frac{\beta_0}{4\pi}
\left \{1+\frac{\beta_1}{2\beta_0}\alpha_s + \rho_2\alpha_s^2+\dots \right \}
\label{RGE}
\end{xalignat}
The $\beta$ function describes the loop corrections to the
gluon propagator, to leading order as:
\vspace*{-.4cm}
\begin{xalignat}{3}
\beta_0&=\frac{11}{3}C_A-\frac{2}{3}n_f &\quad
\setlength{\unitlength}{1.mm}
\begin{fmffile}{mf_betaloop1}
\parbox[c]{35mm}{
   \begin{fmfgraph}(30,10) \fmfleft{i} \fmfright{o1}
    \fmfpen{thin} 
    \fmf{gluon}{i,v1}   
    \fmf{gluon,left=0.8,tension=.6}{v1,v2,v1}
    \fmf{gluon}{v2,o1}
   \end{fmfgraph}
   }
\quad\quad
\parbox[c]{35mm}{
   \begin{fmfgraph}(30,10) \fmfleft{i} \fmfright{o1}
    \fmfpen{thin}
    \fmf{gluon}{i,v1}   
    \fmf{plain,left=0.6,tension=0.4}{v1,v2,v1}
    \fmf{gluon}{v2,o1}
   \end{fmfgraph}
   }   
\end{fmffile}
\end{xalignat} 
It is influenced by all strongly interacting particles.
A measurement of the $\beta$ function implies model independent limits on
hypothetical strongly interacting particles.
Such a measurement has recently been performed \cite{ralle} based on the 
Renormalisation Group Invariant (or Improved)  pert.~theory (RGI p.t.). 
It applies to observables $R\propto\langle y \rangle/A$ 
depending on a single energy scale which are normalised 
such that their pert.~expansion starts with \as. 
RGI p.t. requires the observables to fulfil the RGE.
This equation is solved as for \as\ yielding an $R$ dependent
integration constant, $\Lambda_R$, which can be determined from a fit to the 
data.
$\Lambda_R$ is exactly related to $\Lambda_{\overline{MS}}^{QCD}$.
RGI p.t. is numerically equivalent to the theory in the Effective
Charge (ECH) scheme. E.g. RGI determines $\Lambda_{\overline{MS}}^{QCD}$ 
without freedom of renormalisation scheme choice. It is important to note that
RGI p.t. implicitely resums UV divergencies. Power corrections can be included.

The RGI prediction has been fitted to the energy dependence 
of 7 event shape means \cite{ralle}.
The description of the data is equally good as with  PC's  
though with less parameters. For RGI the power
terms turned out to be compatible with zero ($\lesssim 2\%$ at the $Z$) 
and in contrary to the $\overline{MS}+$PC case
all observables can be described with a single value of
$\alpha_s=0.1201\pm0.0020$.
I.e. the energy dependence of mean values can be described at the 2\% level
without hadronisation correction and without renormalisation scale freedom. 
This implies that in the $\overline{MS}$ scheme power terms
for mean values 
to a large part parameterise missing higher order terms. In fact it is even 
possible to predict the ``n.p.'' terms using  RGI \cite{ralle}.

To leading order the RGE (\eq{RGE}) specifies a straight line when plotting
the inverse of the observable vs. the logarithm of the energy, $\ln Q$. 
Thus, up to a small correction the slope of the data directly determines 
$\beta_0$. In order to optimally determine the $\beta$ function  
data on $\langle 1-T \rangle$
has been chosen spanning the energy range 
15 to 205\GeV (\fig{betaplot}). A linear fit yields:
\begin{xalignat}{1}
\frac{d \langle 1-T \rangle^{-1}}{d \ln Q^2} &= 8.70\pm 0.35 ~~~,
\end{xalignat}
consistent with the expected QCD value $8.32$. 
This result model independently 
excludes supersymmetric gluinos  up to masses of
30--40\GeV\ \cite{ralle}.
Assuming QCD $\beta_0$ or $n_f$ can be alternatively determined:
\begin{xalignat}{3}
\beta_0&=7.83 \pm 0.32\qquad & n_f&=4.75 \pm 0.44~~~.
\end{xalignat}
This result is a clear measurement 
independent of assumptions about the $\beta$ function unlike
the case when using models in \as\ determinations.
It is superior to results from  \lep\ 
($\beta_0=7.67\pm1.63$) \cite{lepwgcomb} or 
world data on \as\ ($\beta_0=7.76\pm0.44$) \cite{bethke}.

\section{Summary}
\wi 0.49\textwidth
\fwi 0.99\wi
\gwi 0.99\wi
\begin{floatingfigure}[bh]{\wi}
\flushleft
\begin{minipage}[t]{\wi}
\includegraphics[width=\fwi,height=\gwi]{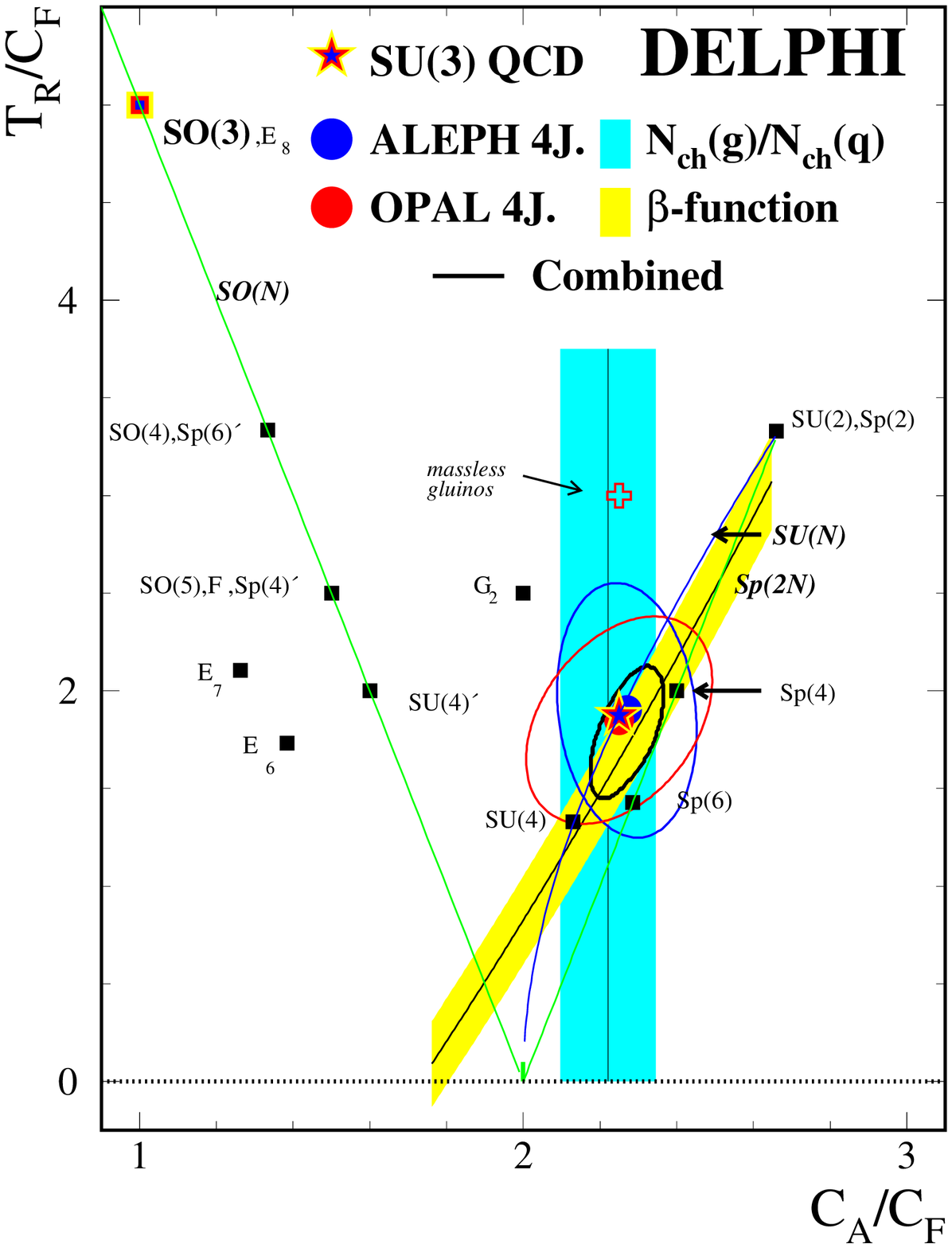}
\caption{\it Casimir factor ratios for several groups compared with
constraints from various measurements.}
\label{wim}
\end{minipage}
\end{floatingfigure}
\noindent 
Measurements performed at \ee\ machines, especially those at and
above the $Z$ pole over the past decade provide a vast amount of hadonic 
physics results. For this talk the selected results appeared 
during the last few years. 

Understanding of the dynamical behaviour of hadrons has been
attained. Most important ingredience of the description here is
the coherence of gluon radiation. Moreover it has recently been possible to
verify the different coupling strengths of quark gluon and gluon gluon vertices
directly from the hadrons created in three jet events.

The instrumentation of the experiments allowed
detailed measurements of heavy quark production. Measurements of the heavy
quark fragmentation function will serve as an important input to
hadron colliders. The measurement of the running of the $b$ quark mass 
tests the loop structure of the $b$ quark propagator.

Several types of \as\ measure-- ments performed lead to a consistent result for
 the
strong coupling, though with a still sizable uncertainty at the few $\%$ 
level. In
most cases the uncertainty is dominated by the theory uncertainty. For a
sizable reduction in \as\ error improved predictions are required.

The measurements of the colour factors from 4 jet events, the multiplicity
ratio obtained in gluon and quark jets and the quantum loops verified
by the measurement of the $\beta$ function strongly 
restrict the gauge  group of
strong interactions  to $SU(3)$ ~/~ QCD (\fig{wim}).\\

\section*{Acknowledgement}
I thank 
P. Abreu, 
D. Duchesneau, 
G. Moneti, 
D. Muller, 
O. Passon, 
M. Siebel,
H. Stenzel, 
R. Seuster 
and D. Wicke for providing input to this talk and
O.P. and M.O. Dima for reading the manuscript.


\end{document}